\documentclass{aastex}
\usepackage{spr-astr-addons}
\usepackage{url}

\RequirePackage{color}

\newcommand{\galex}{{\em GALEX}}
\newcommand{\teff}{$T_{\mathrm{eff}}$}
\newcommand{\mteff}{T_{\mathrm{eff}}}

\newcommand{\logg}{$\log{g}$}
\newcommand{\mlogg}{\log{g}}

\newcommand{\metal}{[M/H]}
\newcommand{\mmetal}{\mathrm{[M/H]}}

\newcommand{\afe}{[$\alpha$/Fe]}

\begin{document}

\title{Statistical properties of the \galex\ spectroscopic stellar sample}
\slugcomment{To appear in UV Universe special issue}
%% Running heads
\shorttitle{GALEX spectroscopy}
\shortauthors{Bertone and Chavez}

\author{Emanuele Bertone} \and \author{Miguel Chavez}
\affil{Instituto Nacional de Astrof{\'\i}sica, \'Optica y Electr\'onica, Luis
  E. Erro 1, 72840 Tonantzintla, Puebla, Mexico}
\email{ebertone@inaoep.mx}

\begin{abstract}
The \galex\ General Data Release 4/5 includes 174 spectroscopic tiles, obtained
from slitless grism observations, for a total of more than 60,000 ultraviolet
spectra. We have determined statistical properties of the sample of \galex\
stars. We have defined a suitable system of spectroscopic indices, which
measure the main mid-UV features at the \galex\ low spectral resolution and we have
employed it to determine the atmospheric parameters of stars in the range
$4500 \lesssim \mteff \lesssim 9000$~K. Our
preliminary results indicate that the majority of the sample is formed by main
sequence F- and G-type stars, with metallicity $\mmetal \gtrsim -1$~dex.
\end{abstract}


\section*{Introduction}
\label{sec:intro}
The ultraviolet (UV) radiation from space is absorbed by the Earth atmosphere,
thus requiring observations from space-borne observatories. 
At present, the only fully UV-dedicated space mission is the {\it Galaxy
  Evolution Explorer} (\galex), which has been observing the UV sky
since 2003. Up to now, it has provided photometry of more than 200 million
objects (see Bianchi, this volume), nevertheless its spectroscopic database is still almost completely
unexploited. To our knowledge, just very few works have been published which
made use of \galex\ spectra \citep{bello09,rosa09,barger10,cowie10,feldman10}.
In this work, we conduct a general analysis of the
whole spectroscopic database aimed at identifying and characterizing the
variety of observed stars, which could be later applied to studies of Galactic
and extragalactic stellar populations.
In particular, we define suitable spectral tools to determine statistical
properties of the stellar sample, through their main atmospheric parameters.

\section{The \galex\ spectroscopic database.}

\galex\ is a NASA small explorer mission \citep{martin05} whose
  main goal is to carry out a two-band 
UV photometric all-sky survey. Being equipped with a grism,
it can also provide slitless spectroscopy of the entire field-of-view (1.2~deg) in the Far-UV (FUV; 1300--1820~\AA) and Near-UV (NUV; 1820--3004~\AA),
at resolutions of $R=\lambda/\Delta\lambda \sim 200$ 
and $R \sim 100$, respectively.
The \galex\ General Release 4/5\footnote{http://galex.stsci.edu/GR4/} contains 174 spectroscopic tiles for a total
observing time of about 2.5~Msec. The database includes the spectra of more than 60000 objects.
For extracting spectra, we have used the automatic pipeline\footnote{ http://www.galex.caltech.edu/DATA/gr1\_docs/
GR1\_Pipeline\_and\_advanced\_data\_description\_v2.htm }, which provides the
wavelength calibrated and flux calibrated monodimensional spectra of the
objects in each tile.

\section{Flux calibration: \galex~vs. IUE}

\begin{figure}[t]
\centering
\includegraphics{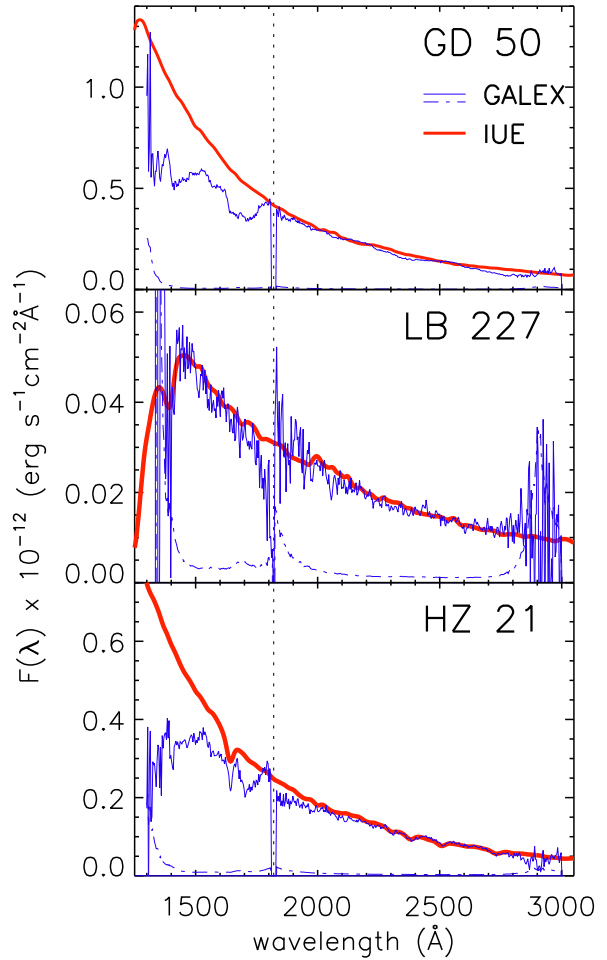}
\caption{IUE (Holberg et al. 2002; thick red line) and \galex\ GR4/5 (thin
  blue line) spectra of the white dwarfs GD50 and HZ21. The \galex\ flux error
  is also shown with a dash-dotted line.}
\label{fig:wd}
\end{figure}

A reliable flux calibration is undoubtedly a crucial point whenever colors are
considered or large wavelength intervals are used to extract
stellar atmospheric parameters.
The \galex\ flux calibration \citep{morrissey05,morrissey07} have been
obtained from a series of observations of white dwarfs (WD) taken from the
catalog of \cite{bohlin01}. To verify the robustness of
this calibration, we have compared the WDs in the \galex\
spectroscopic database which are in common with the study of \cite{holberg03},
who provide a series of high quality IUE spectra of Galactic white dwarfs. In
Fig.~\ref{fig:wd}, we plot the UV spectral energy distribution (SED) of three
objects (GD~50, LB~227, and HZ~21) that show that the best agreement between
IUE and \galex\ data is achieved in the NUV interval, while the \galex\ FUV
range presents severe flux calibration problems, which consist in a
overall underestimation of flux and the appearance of spurious
features. We therefore limited our present study to NUV wavelengths between
2000 and 2830~\AA, as, outside this interval, the \galex\ NUV detector
sensitivity decreases drastically. 
We would like to mention, however, that the ``blue'' (i.e., short-wavelength) edge of \galex\ sensitivity in
the NUV drops at shorter wavelengths than IUE, allowing a better spectral
morphological analysis over the spectral window 2000--2300~\AA.

\section{The stellar sample.}

Each spectroscopic tile has also been observed in imaging mode, so that it is
possible to assign each spectrum to its corresponding source in the NUV image.
We made use, for each of the 174 spectroscopic tiles, of the
data present in the photometric catalog file
(*-xd-mcat.fits) and the spectra provided by the \galex\ automatic pipeline (*-xg-gsp.fits).
Since the goal of this work is to assess the potential of \galex\ data to
determine the main properties of stars, we extracted point-like objects from the entire database by means of the following criteria:

\begin{itemize}
\item ellipticity$\leqslant$0.2 in the NUV image;
\item (SExtractor parameter) class\_star$\geqslant$0.8, in the NUV image, and
\item a median SNR/px$\geqslant$5 in the NUV spectral interval. This criterion
  by itself decreases the number of spectra to about 11000.
\end{itemize}

\begin{figure}
\centering
\includegraphics[width=5.2cm]{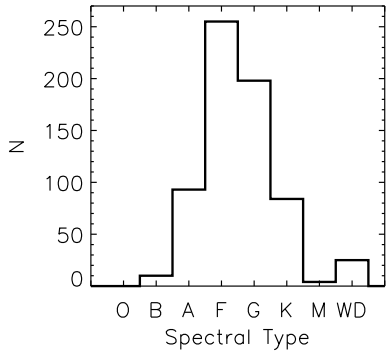}
\caption{Spectral type distribution of the subsample of SIMBAD stars.}
\label{fig:histosimbad}
\end{figure}

The selected analysis sample is formed by a total of 6037 spectra.

\subsection{Sample properties and non-stellar contamination.}

The adopted selection criteria do not exclude the presence in the
sample of other kind of sources, besides stars. Therefore, in order to have an indication on the amount of contamination, we proceeded to
cross-correlate the position of all sources in our sample with the SIMBAD
database\footnote{http://simbad.u-strasbg.fr/simbad/}.
We found a total of 2132 objects, which is about 35\% of our total
sample. They include 235 entries, which are classified as non-stellar (QSOs,
galaxies, and so on), and 118 objects which are only identified as being
sources emitting in UV, X, IR, submm, etc.
We concluded that the expected fraction of contamination from
non-stellar (or unclassified) objects of our sample should be around 16\%,
which imply a total number of stars of about 5000.

\begin{figure}[!b]
\centering
\includegraphics[width=84mm]{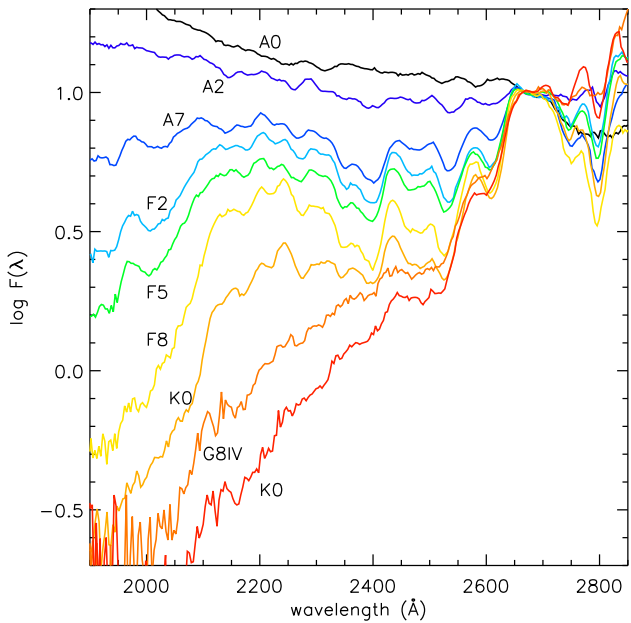}
\caption{Selected \galex\ spectra of stars along the spectral type sequence.}
\label{fig:spectra}
\end{figure}

The distribution of the spectral types of the stars present in our sample is
shown in Fig.~\ref{fig:histosimbad}: the peak corresponds to F and G types,
with a significant contribution of A and K type stars. In an effective
temperature scale, this translates to roughly the interval 4000--10000~K.
In this plot we incorporate those stellar objects that have a spectral
classification in SIMBAD.
The fraction of white dwarfs with respect of all other stars can be biased
from the fact that they were selected for calibration purposes.
Examples of high SNR \galex\ stellar spectra along the spectral type sequence
are shown in Fig.~\ref{fig:spectra}.

\begin{figure}[!b]
\centering
\includegraphics[width=6cm]{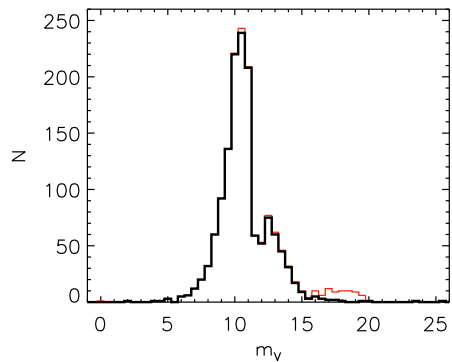}
\caption{Histogram of the apparent V magnitude of the sample of objects with
  \galex\ spectra
  present in the SIMBAD database. The thin red line represents the whole
  sample, while the thick black line indicates the distribution of stellar
  objects.}
\label{fig:histomag}
\end{figure}

The apparent V magnitude distribution (Fig.~\ref{fig:histomag}) of the SIMBAD sample is highly peaked at
around 9--11~mag. The distribution of stellar objects (thick black line) almost
coincide with the total distribution up to $\sim$15~mag: this implies that
they mostly belong to the Galactic thin and thick disk
components\footnote{A typical G dwarf of $M_V=+5$~mag would have a distance of
 158~pc, if $m_V=11$~mag, and 1000~pc, if $m_V=15$~mag.}
\citep[e.g.,][]{juric08}. Point-like non-stellar objects are the
majority of dimmer objects in the 15--20~mag interval.

\section{A new set of spectroscopic indices for \galex\ observations.}

\begin{table}[!t]
\centering
\caption{Index passband definition.\label{tab:def}}
\begin{tabular}{@{}lccc@{}}
\tableline
Index name$^{a}$          & Blue band  & Index band & Red Band \\
\tableline                                       
gFe\textsc{i}2260    & 2230--2256 & 2259--2285 & 2294--2320 \\
gFe\textsc{ii}2332   & 2294--2320 & 2338--2364 & 2421--2447 \\
gFe\textsc{ii}2402   & 2294--2320 & 2380--2406 & 2421--2447 \\
gBL2538              & 2448--2474 & 2511--2537 & 2644--2670 \\
gFe\textsc{ii}2609   & 2448--2474 & 2597--2623 & 2644--2670 \\
gBL2730              & 2732--2758 &            & 2674--2703 \\
gMg\textsc{ii}2800   & 2784--2810 &            & 2674--2703 \\
g2609/2680           & 2597--2623 &            & 2674--2703 \\
g2000/2180           & 2000--2026 &            & 2156--2182 \\
\tableline
\multicolumn{4}{l}{\footnotesize $^{a}$The g prefix is added to differentiate these indices} \\
\multicolumn{4}{l}{\footnotesize from those defined in the original Fanelli's (IUE) system.}
\end{tabular}
\end{table}

The use of spectrophotometric indices to extract selected information from spectra has been a very common practice in the optical; for
example, the Lick/IDS system \citep[and references therein]{trager98} has
found a great number of applications in many different research fields.
In the UV, however, the use of this kind of analytical tool has been much less
widespread. Recently, \cite{chavez07,chavez09} have studied the mid-UV
spectral morphology of stars and evolved stellar populations by means of the
\cite{fanelli92} set of indices computed in IUE observations.

For the analysis of our sample, we used a set of 9 spectrophotometric
indices carefully defined to be suitable for the \galex\
spectral resolution and to maximize the sensitivity to stellar parameter
variations. Since the indices mainly depend on the strength of absorption
features, they are most suitable for late-type stars, while they decrease
their effectiveness for hot O--B stars.

\begin{figure}[!t]
\centering
\includegraphics[width=8cm]{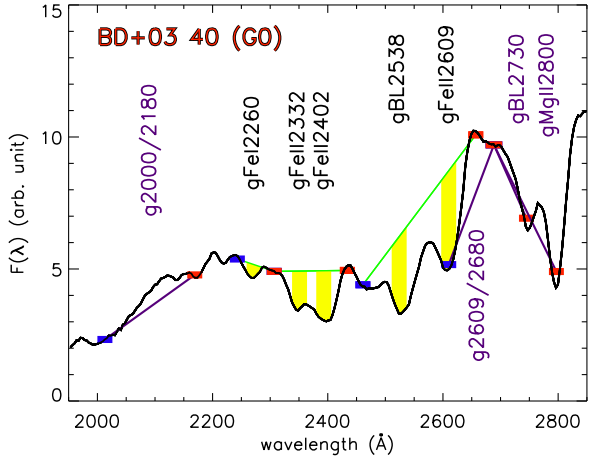}
\caption{The \galex\ spectrum of the star BD+03~40 along with the passbands
 of the set of spectroscopic indices.}
\label{fig:indexdef}
\end{figure}

The index ID and passbands are reported in Table~\ref{tab:def} and shown in Fig.~\ref{fig:indexdef}.
For 5 indices we have adopted the definition given by Eq.~5 in \cite{fanelli92}. In this definition the flux in the index band is 
compared to a pseudo-continuum, obtained by the interpolation of the fluxes in
the two side bands; the other 4 indices are narrow-band colors, which
measure the difference in magnitude between the blue and red band. Note that,
due to the red edge cut-off sensitivity, it was not possible to define a
suitable red band for the indices measuring the strength of the Mg~\textsc{ii} doublet 
at 2800~\AA\ and the blend around 2730~\AA: we have therefore defined two colors, where 
the blue and red bands are inverted; in this way, the indices have increasing values with 
increasing line absorption.
The first 8 indices of the table measure the intensity of the same spectral
features of the Fanelli et al. system, apart from
gBL2730, which merges together the two IUE absorption blends at 2720 and
2740~\AA\ due to the lower \galex\ spectral resolution. 
The brand new index that we have defined, the color g2000/2180, takes
advantage of the higher sensitivity of \galex\ at $\lambda \lesssim 2300$~\AA\
with respect to IUE. This property also allows more precise measurements of
the gFe\textsc{ii}2332 and gFe\textsc{ii}2402 indices, which \cite{chavez09}
point out as being good tracers of relatively young and metal rich stellar populations.

The new indices were measured in three different spectral samples:
\begin{itemize}
\item the 6037 selected \galex\ point-like sources;
\item the ATLAS library of synthetic stellar SEDs of
  \cite{castelli03}\footnote{http://wwwuser.oat.ts.astro.it/castelli/}, after
  having homogenized them to the same spectral resolution and wavelength
  sampling of the \galex\ data. Since the flux is sampled at
  $\Delta\lambda=10$~\AA, this library is suitable for analysing \galex\
  spectra;
\item a chosen sample of IUE stars, from the \cite{wu83} catalog, whose
  atmospheric parameters are known. In this case, we have also properly degraded the
  original IUE spectra to the \galex\ resolution. This set of objects was used in
  the calibration process described in the following Sec.~\ref{sec:calib}.
\end{itemize}

\begin{table}[!t]
\begin{center}
\caption{Linear transformation parameters.\label{tab:calib}}
\begin{tabular}{lrrrrr}
\tableline
Index & a & b & $\sigma$ & used & excl. \\
\tableline
gFe\textsc{i}2260  &  0.016    &  0.597   &  0.022   &    71     &  23 \\
gFe\textsc{ii}2332 &  0.029    &  1.369   &  0.128   &    79     &  15 \\
gFe\textsc{ii}2402 &  0.062    &  1.699   &  0.177   &    82     &  13 \\
gBL2538            &  0.025    &  1.419   &  0.095   &    61     &  34 \\
gFe\textsc{ii}2609 & -0.008    &  1.285   &  0.097   &    67     &  28 \\
gBL2730            &  0.023    &  1.430   &  0.127   &    79     &  15 \\
gMg\textsc{ii}2800 &  0.223    &  1.133   &  0.274   &    84     &  10 \\
g2609/2680         &  0.019    &  1.151   &  0.088   &    72     &  23 \\
g2000/2180         &  0.125    &  1.365   &  0.426   &    80     &  13 \\
\tableline
\end{tabular}
\end{center}
\end{table}

\begin{figure*}
\centering
\begin{tabular}{ccc}
\includegraphics[scale=0.43]{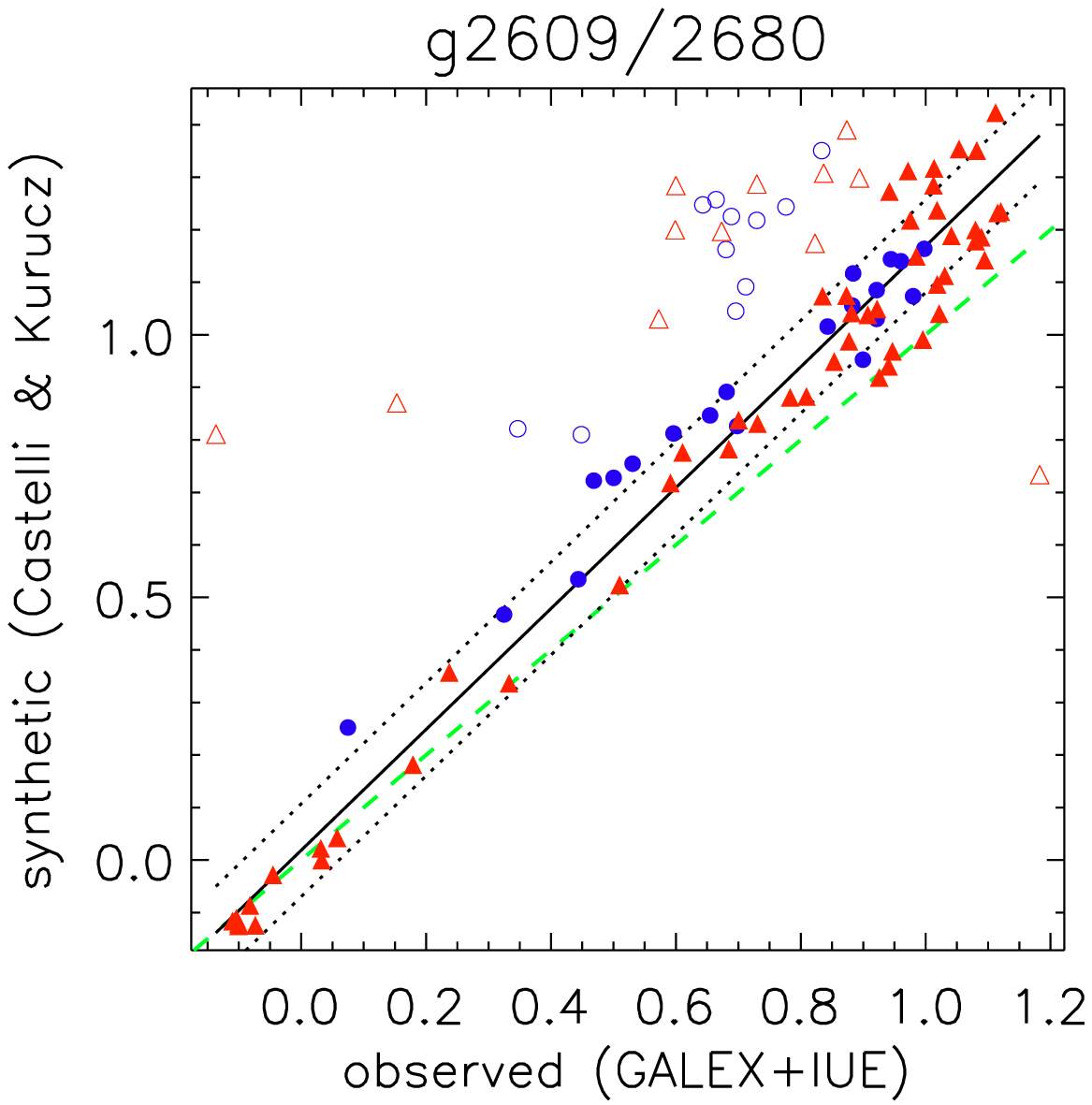}&
\includegraphics[scale=0.43]{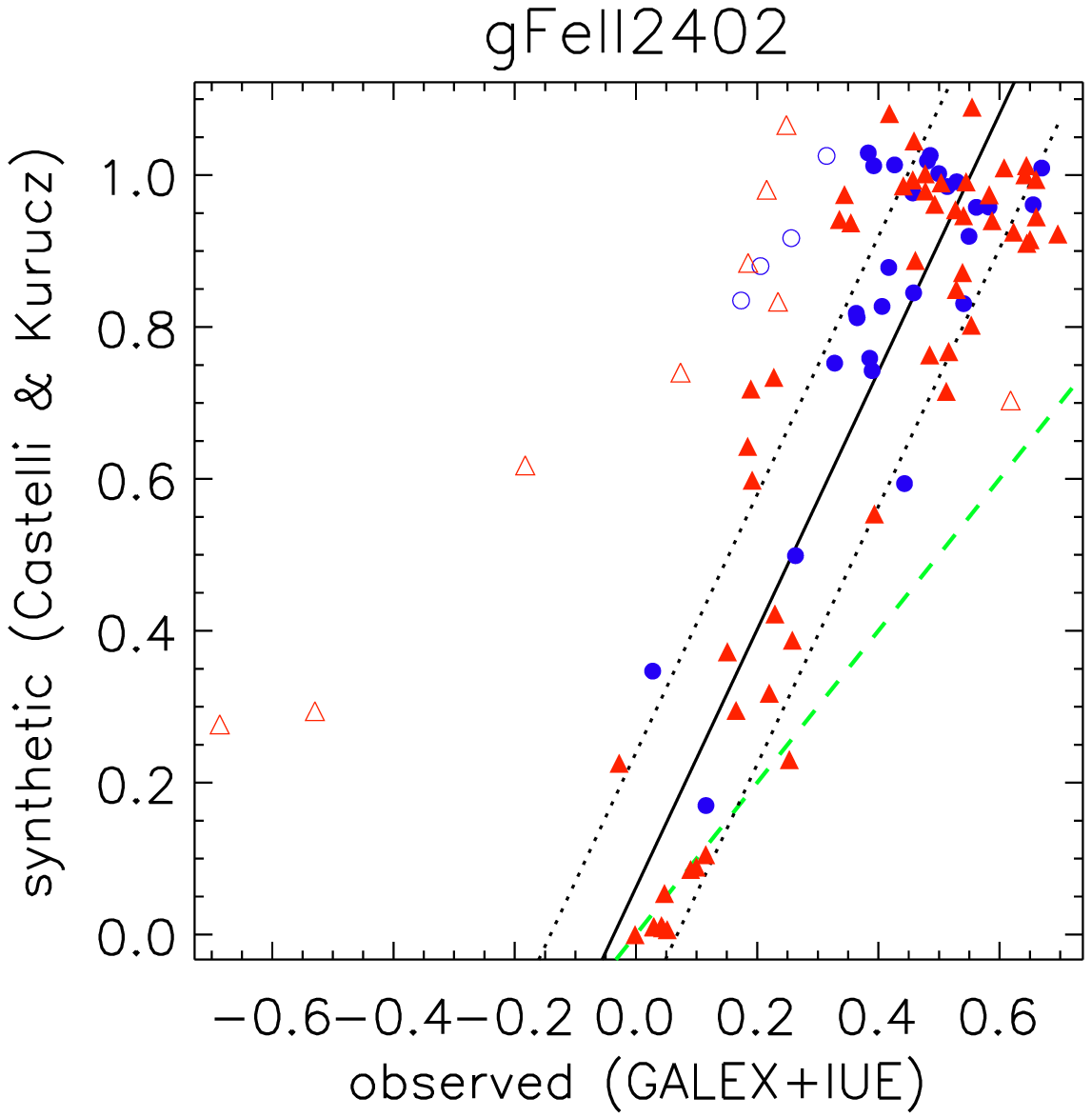}&
\includegraphics[scale=0.43]{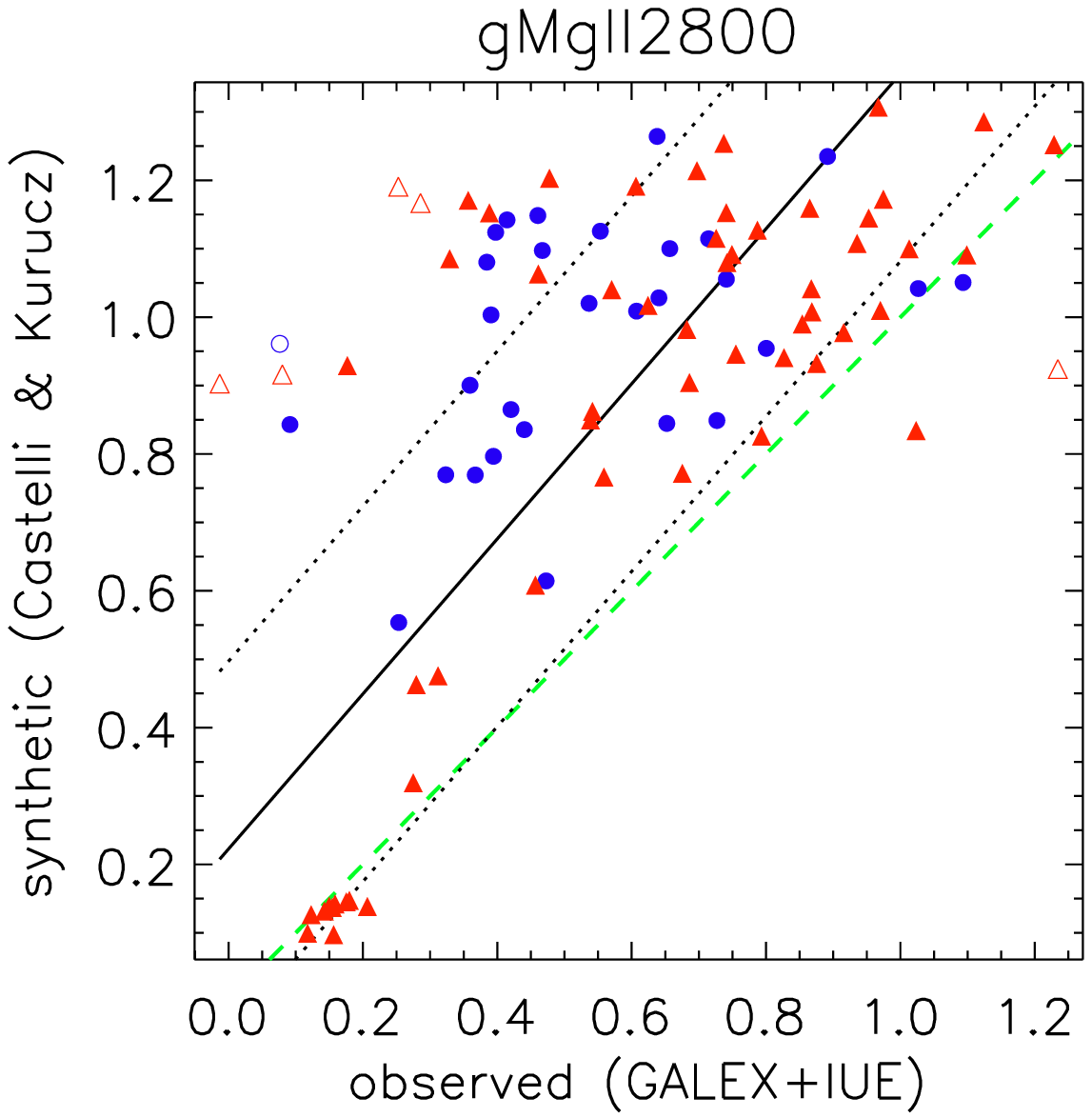}\\
\end{tabular}
\caption{Theoretical vs. observed indices for the \galex\ (blue circles) and
  IUE (red triangles) samples. The solid black lines indicates the best linear
  fit, which did not take into account the outliers (open symbols); the
  parallel  dotted lines lie 1$\sigma$ apart. The green dashed line shows the
  one-to-one correlation.}
\label{fig:calib}
\end{figure*}

\begin{figure*}
\centering
\includegraphics{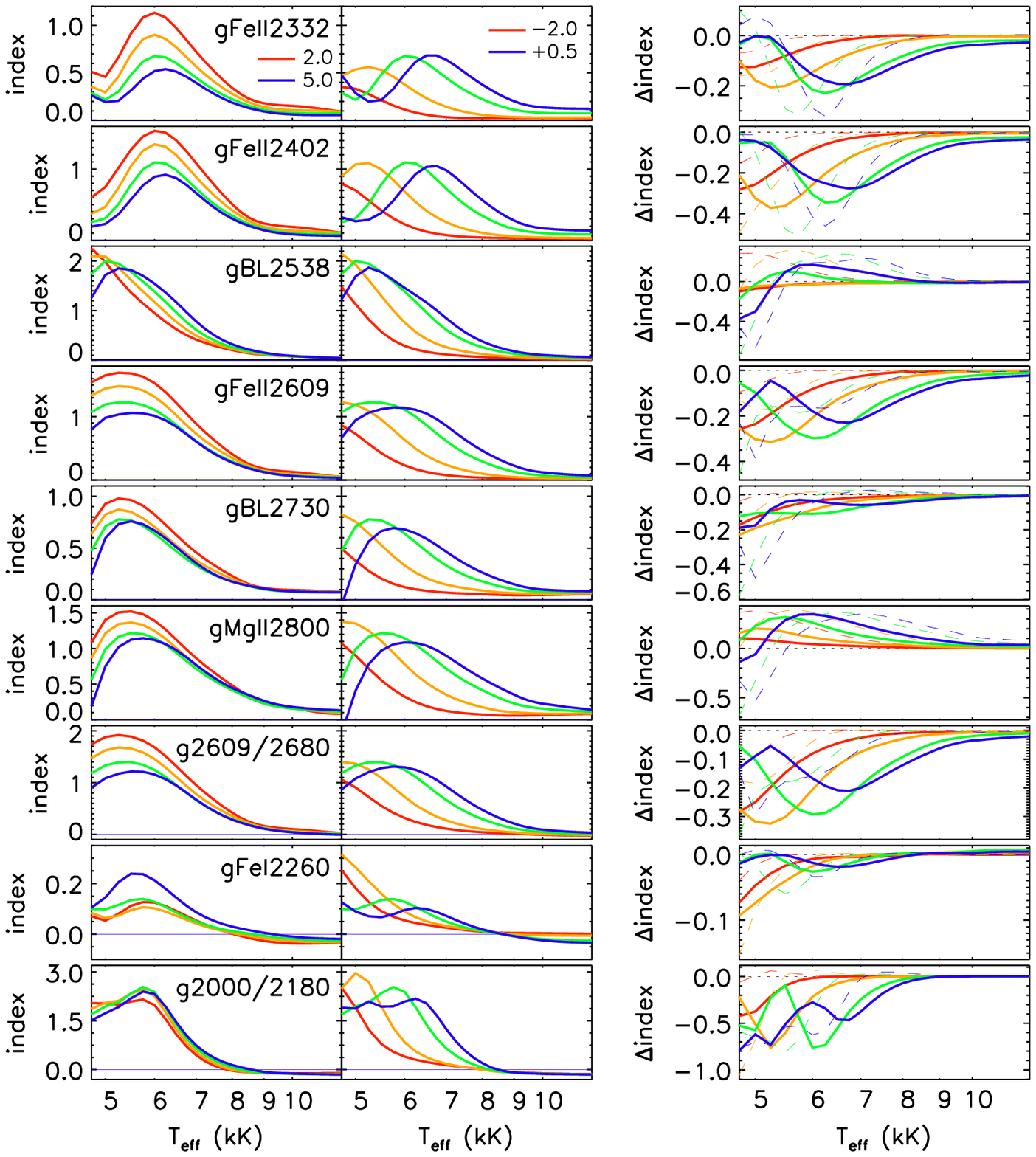}
\caption{{\it Left column:} Theoretical indices, at fixed \metal=0.0, as a function of effective
  temperature for different values of \logg= 2 (red curve), 3
  (yellow), 4 (green), and 5~dex (blue). {\it Central column:}
  Theoretical indices, at fixed \logg=4~dex, as a function of effective
  temperature for different values of solar-scaled metallicity [M/H]= -2
  (red curve), -1 (yellow), 0 (green), and +0.5 (blue). {\it Right
    column:} Effect of the $\alpha$-enhancement on the theoretical indices for
  the same metallicity values shown in the central column: the thick curves
  show the [$\alpha$/Fe]=+0.4 minus the solar-scaled indices once the
  abundances are re-scaled to have the same $Z$ as the solar-scaled models,
  while the thin dashed lines show the same difference obtained by adding
  0.4~dex to the number density of $\alpha$ elements.}
\label{fig:param}
\end{figure*}

\subsection{Calibration of the theoretical indices.}
\label{sec:calib}

Since the ATLAS SEDs do not perfectly reproduce the spectra of real stars
\citep{bertone04,bertone08,rodr05}, the theoretical indices must be converted
to the observed system, before applying them to the analysis of the \galex\
sample. We performed this calibration, as in \cite{chavez07}, by means of a
sample of stars, which were observed by \galex\ and whose complete set of
atmospheric parameters (\teff,\logg,\metal) is given in the PASTEL catalog
\citep{soubiran10}. Since the total number of these stars is not large enough
to provide a good calibration, we added a sample of IUE star from the
\cite{wu83} catalog. The theoretical spectra, with parameters corresponding to
each of the calibration stars, were obtained by a tri-linear interpolation of
the \cite{castelli03} grid. 
Figure~\ref{fig:calib} shows the comparison between the observed and the
theoretical indices for three indices, which are representative of
the different degrees of agreement, from one of the best cases (g2609/2680) to
the worst (gMg\textsc{ii}2800). The trasformation equations are
obtained by an iterative process of linear interpolation (${\rm index}_{\rm
  theor} = a + b \times {\rm index}_{\rm obs}$), excluding at each step the
outliers which stand more than 2$\sigma$ away from the fitted value.
In Tab.~\ref{tab:calib} we report, along with the $a$,$b$ parameters, the
final rms and the number of stars used and excluded in the fitting process.
There is a clear tendency of the theoretical value to be higher than the
observed data, the only exception being the gFe\textsc{i}2260 index.
This disagreement makes the tranformation process compulsory if the goal is to
apply this set of theoretical indices to the determination of the atmospheric
parameters of stars.

\subsection{Theoretical indices vs.~stellar atmospheric parameters.}

Within the theoretical framework, we explored the sensitivity of each
index with respect to the fundamental stellar atmospheric parameters:
effective temperature (\teff), surface gravity (\logg), global metallicity (\metal), and the \afe\ enhancement. The results are displayed in
Fig.~\ref{fig:param}.
The indices that measure common spectral features with those of \cite{fanelli92} show similar trends, with respect to \teff, \logg, and \metal, as those
calculated with the Fanelli's passbands definition \citep{rodr04,chavez07}. Apart from the two bluest indices, the values are larger for lower
gravities. The slope g2000/2180 is an interesting case, since it is
almost independent of gravity over the whole temperature range. As far as
\metal\ is concerned, all indices show a strong dependence and a similar
pattern: the temperature at the peak increases with increasing metal
abundance. 

\cite{chavez09} found a strong effect of the
$\alpha$-enhancement in the SEDs of stars and simple stellar populations all
over the mid-UV interval. In what follows, we explore its impact on the
spectroscopic indices. In the \cite{castelli03} library, the $\alpha$-enhanced models and
fluxes were computed by adding an extra abundance of +0.4~dex to each $\alpha$
element, which causes an increase of about +0.3~dex to the global metallicity
$Z$ with respect to the nominal one. Therefore, to compare the results at
equal $Z$, we scaled down all abundances of the \afe=+0.4 models to reach the 
same global metallicity as the solar-scaled models and we computed the
correspondent SEDs by linear interpolation in the grid of theoretical fluxes.
The abundance rescaling produces a significant reduction of the non-$\alpha$
elements, in particular of the Fe, which is reflected in a decrement of almost
all the \galex\ indices, whose intensity is strongly dependent from the
absorption of iron atoms, which is one of the main sources of opacity in the
mid-UV. The indices gBL2538 and gMg\textsc{ii}2800 are the only ones which present
non negligible positive differences in $\alpha$-enhanced spectra: this is due,
in both cases, to the presence of magnesium and, subordinately, titanium
absorption lines inside the central index band.
In Fig.~\ref{fig:param}, we also trace a comparison of the indices computed
for the original \cite{castelli03} \afe=+0.4 SEDs and the
correspondent solar-scaled ones: in this case, besides the effect of the
enhancement of the $\alpha$ elements, the difference takes into account the
global increasing in metallicity; therefore, the difference has higher values
(i.e., less negative) at higher temperatures, where the index strength
increases with increasing global metallicity, and, viceversa, $\alpha$-enhanced
indices are lower were the index dependence to metallicity is reversed.

We conducted a complementary analysis assessing the effect of
interstellar extinction on the indices. For this test we considered
$E(B-V)$=0.1, which is a typical value for the objects of our sample. The
results indicate that index variations due to extinction are negligible.

\begin{figure}[!t]
\centering
\begin{tabular}{c}
\includegraphics[scale=0.52]{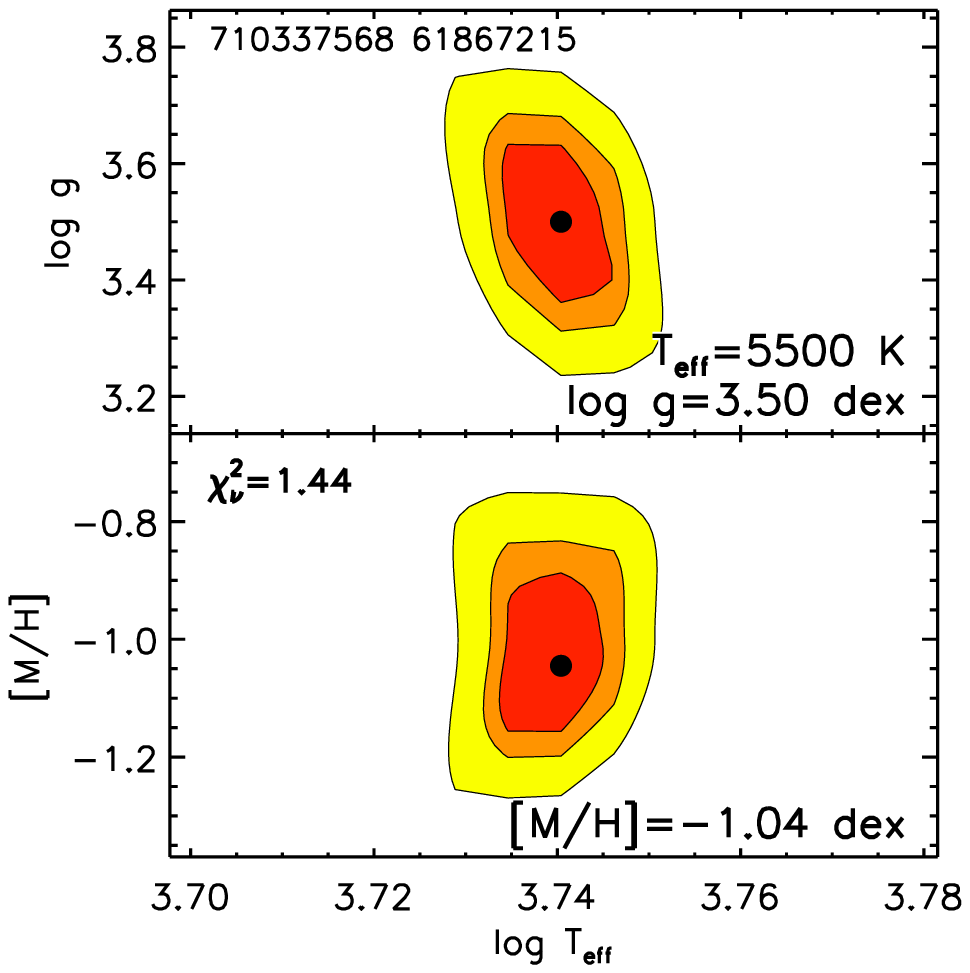}\\
\includegraphics[scale=0.52]{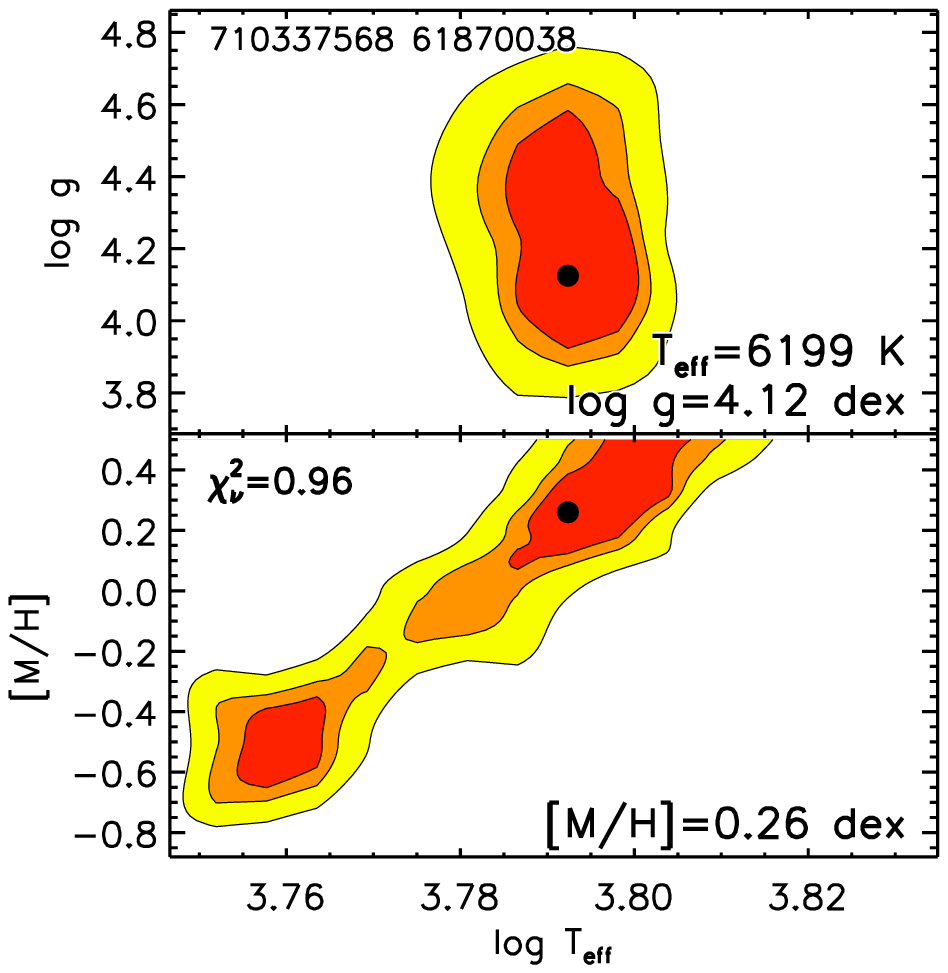}\\
\includegraphics[scale=0.52]{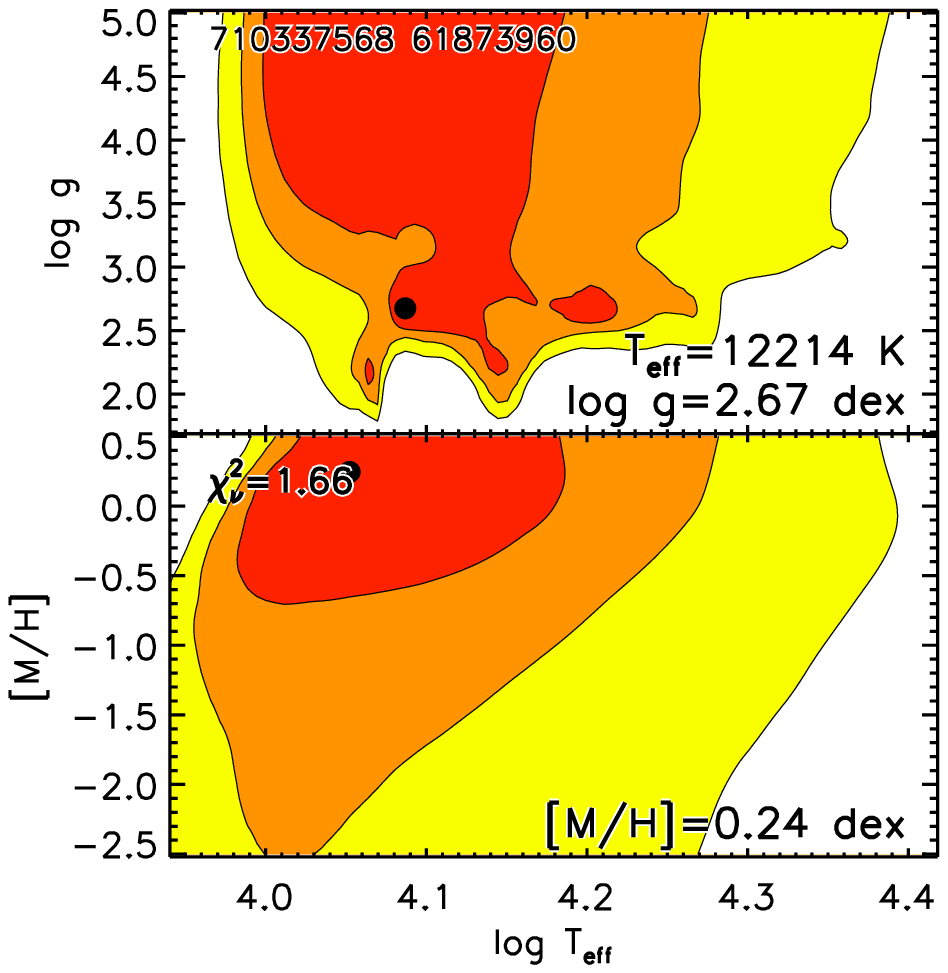}\\
\end{tabular}
\caption{$\chi^2_{\nu}$ contour plots in the \logg~vs.~\teff\ plane (upper
  panels) and \metal~vs.~\teff\ plane (lower panels) for 3 \galex\ objects. The
  filled circles mark the loci of the minimum $\chi^2_{\nu}$, while the grey
  surfaces bounded by the 1-, 2-, and 3-$\sigma$ error contours. In
  the panels are also indicated the \galex\ identification code, the best
  values of the atmospheric parameters, and of the minimum $\chi^2_{\nu}$.} 
\label{fig:contour}
\end{figure}

\section{Atmospheric parameters of the \galex\ objects.}

We have used the set of indices defined here to determine the three main
stellar atmospheric parameters for all the objects of
the selected sample. For non-stellar objects contaminating the sample the
results are, of course, meaningless. In this first work on \galex\ spectra, we
restricted the examination of the results to a statistical analysis. 

We obtained effective temperature, surface gravity, and global metallicity of
each object by minimizing the chi-square distances between the observed set of
indices and the theoretical ones. The reduced chi-square is defined as:
\begin{equation}
\chi^2_{\nu, j} = \frac{1}{\nu} \Sigma_{i} \frac{(I_{\rm theor, i, j} - I_{\rm obs,
    i})^2}{\sigma^2_{\rm  obs, i}}\, .
\end{equation}
where $i$ runs over the spectroscopic indices $I$, while $j$ indicates the
theoretical model and $\nu$ is the number of indices which could be computed
from the spectrum. The error $\sigma$ for each observed index is determined via a Montecarlo
  computation: a random error taken from a Gaussian distribution, with
  standard deviation equal to the observed flux error, is added to the original \galex\ spectrum and
the indices are computed on this new spectrum. This is repeated 200 times and
the error of each index is given by the standard deviation of the index
distribution.

We searched for the combination of \teff, \logg, and \metal\ that presents
the minimum $\chi^2_{\nu}$, so that we could extract the planes at fixed metallicity and at
fixed gravity that included this point. We then refined the determination of the
minimum on the two surfaces after smoothing them with a quintic surface, by
means of the {\em tri\_surf} IDL function. The loci of the minima allow the
determination of the best values of \teff, \logg, and \metal\ of each object.
In Fig.~\ref{fig:contour}, we present three examples of the contour plots of
the $\chi^2_{\nu}$ in the \logg\ vs.~\teff\ and \metal\ vs.~\teff\ planes: they
show one case where the procedure provides a precise determination of the
stellar parameters (upper panel), a second case where the \teff--\metal\
degeneracy generates two 1-$\sigma$ regions (middle panel), and finally an
example of a failure of the method to determine the parameters with an adequate
precision (lower panel).

\begin{figure}[!t]
\centering
\begin{tabular}{c}
\includegraphics[scale=0.84]{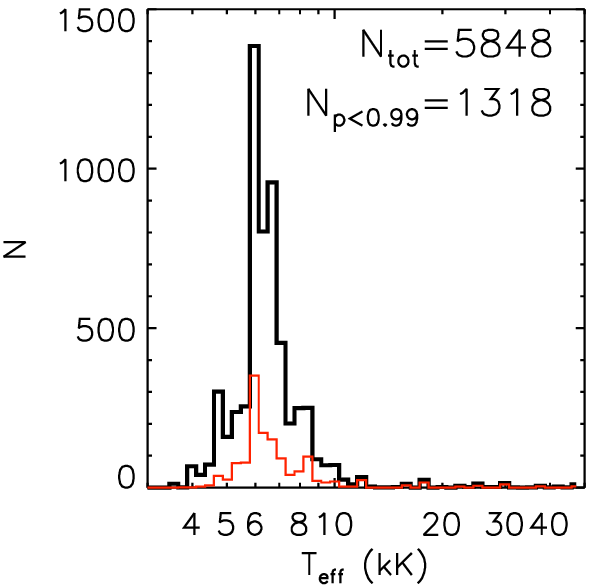}\\
\includegraphics[scale=0.84]{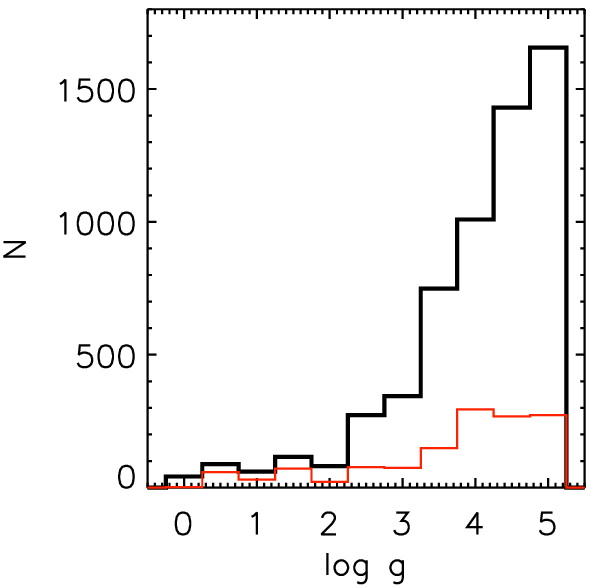}\\
\includegraphics[scale=0.84]{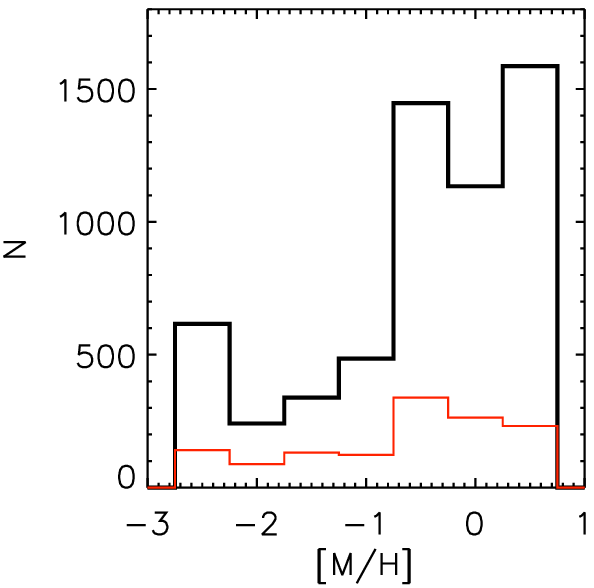}\\
\end{tabular}
\caption{Parameter distributions of the \galex\ sample. The thick black lines
  refer to the whole sample, while the thin red lines show the histograms
  of the results which have a significance level greater than 1\%.}
\label{fig:distr}
\end{figure}

The distributions of the results are shown in Fig.~\ref{fig:distr}. The
histogram of the effective temperature peaks at $\mteff \sim 6000$~K, typical
of early-G or late-F spectral types, and show very large numbers up to $\mteff
\sim 7000$~K (the F-type interval). A significant fraction of objects have
a temperature of later G- and K- (approximately 4000--5800~K) and A-type
($\sim$7000--10000~K) stars, while outside this interval the numbers are very
small. This distribution well agrees with the stellar type
histogram of the SIMBAD stellar sub-sample (Fig.~\ref{fig:histosimbad}). 

The surface gravity distribution indicates a preponderance of main sequence
objects ($\mlogg \geq 4$~dex) with respect to more evolved
stars, while the histogram of the global metallicity shows a majority of stars
with $\metal > -1$~dex.
This is in agreement with the properties of the Galactic stellar populations
of the thin and thick disks \citep[e.g.,][and references therein]{buser99}.

Since the theoretical stellar library includes only two values of
$\alpha$-enhancement, we excluded, for now, the determination of this further parameter. We can say, nevertheless, that, according
 to the behaviour of the indices with \afe\ shown in Fig.~\ref{fig:param}, we expect to have slightly overestimated the
effective temperature and  underestimated the global metallicity if an enhancement is present in some of the stars.

\acknowledgments
The authors acknowledge partial financial support from Mexican CONACyT via
grant 49231-E. This research has made use of the SIMBAD database,
operated at CDS, Strasbourg, France.

\end{document}